\documentclass[preprint,12pt]{elsarticle}

\usepackage{verbatim} 
\usepackage{epsfig}
\usepackage{amssymb} 
\usepackage{mathtools}
\usepackage{lineno}
\usepackage{tikz}        
\usepackage{etoolbox}
\usepackage{ulem}         
\usepackage[caption=false]{subfig}

\begin{document}

\newcommand*{\hwplotB}{\raisebox{3pt}{\tikz{\draw[red,dashed,line 
width=3.2pt](0,0) -- 
(5mm,0);}}}

\newrobustcmd*{\mydiamond}[1]{\tikz{\filldraw[black,fill=#1] (0,0) -- 
(0.1cm,0.2cm) --
(0.2cm,0) -- (0.1cm,-0.2cm);}}

\newrobustcmd*{\mytriangleleft}[1]{\tikz{\filldraw[black,fill=#1] (0,0.15cm) -- 
(-0.3cm,0) -- (0,-0.15cm);}}
\definecolor{Blue}{cmyk}{1.,1.,0,0} 

\begin{frontmatter}

\title{Effects of the body force on the pedestrian and the evacuation dynamics}

\author[add1]{I.M.~Sticco}
 \address[add1]{Departamento de F\'\i sica, Facultad de Ciencias 
Exactas y Naturales, \\ Universidad de Buenos Aires,\\
 Pabell\'on I, Ciudad Universitaria, 1428 Buenos Aires, Argentina.}

 \author[add2]{G.A.~Frank}
 \address[add2]{Unidad de Investigaci\'on y Desarrollo de las 
Ingenier\'\i as, Universidad Tecnol\'ogica Nacional, Facultad Regional Buenos 
Aires, Av. Medrano 951, 1179 Buenos Aires, Argentina.}

\author[add1,add3]{C.O.~Dorso\corref{cor1}}%
 \cortext[cor1]{codorso@df.uba.ar}

 \address[add3]{Instituto de F\'\i sica de Buenos Aires,\\
Pabell\'on I, Ciudad Universitaria, 1428 Buenos Aires, Argentina.}

\begin{abstract}

The Social Force Model (SFM) is a suitable model for describing crowd
behaviors under emotional stress. This research analyzes the role of the body
force in the original SFM. We focused on the parameter associated with the
body stiffness ($k_n$) and its impact on the pedestrian dynamics for two
different geometries: bottlenecks and corridors. Increasing $k_n$ produces
opposite effects on the crowd dynamics for each geometry: an increase of the
crowd velocity  for bottlenecks, and a decrease for corridors.  The former
reflects the fact that an increase in the stiffness reduces the overlap
between pedestrians and, as a consequence, the sliding friction is diminished.
This phenomenon reduces the number of blocking clusters close to the exit
door.  In the case of the corridor, instead, due to the confining walls, the
pedestrians get tight  into a lattice-like configuration due to space
limitations. The friction interaction with the walls determines the velocity
of the whole crowd along the corridor. Additionally, the corridor geometry
generates a flux slowing down for very crowded environments, as observed in
many real-life situations. We also explored the dimensionless parameters that
arose from the reduced-in-units equation of motion and tuned them to reproduce
the qualitative behavior of the empirical fundamental diagram.

\end{abstract}

\begin{keyword}

Pedestrian Dynamics \sep Social Force Model \sep Body Force

\PACS 45.70.Vn \sep 89.65.Lm

\end{keyword}

\end{frontmatter}


\section{\label{introduction}Introduction}

The Social Force Model (SFM) models the crowd dynamics  considering three kinds
of forces: desired forces, social forces, and physical forces. In its original
version, the SFM, addresses two physical forces as essential:   the ``body
force'' and the ``sliding friction''. Both are  inspired by granular
interactions and were claimed to be necessary   for attaining the particular
effects in panicking crowds \cite{helbing_2000}.  The ``sliding friction''
actually proved to be an essential feature of the  ``faster-is-slower'' effect,
although the role of the ``body force'' appears,  at a first instance, not so
clear \cite{dorso_2005,dorso_2007,dorso_2011}. \\

The existence of a ``body force'' in the context of highly dense crowds (say,
more than 5$\,$people/m$^2$) is a commonsense matter
\cite{henein_2007,fruin_1993}. Researchers, however, question the numerical
setting for this force in  the SFM context \cite{lakoba_2005}. As a matter of
fact, the usual   set of parameters provided by  Helbing prevents the excessive
overlapping among pedestrians, but it is known to  accomplish artificially high
force levels \cite{helbing_2000,lakoba_2005,langston_2006,lin_2017}. The force
estimates from the SFM appear to be remarkably higher with respect to the
reported real life data (say, an order of magnitude). The crowd motion
simulations, however, present realistic results
\cite{lakoba_2005,langston_2006,dorso_2017}.  The point seems to be that the SFM
focuses on the ``collective behavior'' due  to clogging, missing the
``individualistic'' perspective  of single pedestrians  or very small groups
\cite{helbing_2000,henein_2007,narain_2009}.  \\

Many researchers realized that modifying the SFM may (partially) fix the
misleading results. It was proposed that the pedestrians'  psychological force
(say, the ``social force'') should be suppressed in the  context of highly dense
crowds  \cite{pelechano_2007,moussaid_2011,alonso_2014,bottinelli_2017}, or
smoothly  quenched according to the crowd density \cite{song_2019}. The authors
in  Refs.~\cite{kabalan_2017,jebrane_2019} further proposed a rigid body model
in  order to completely avoid the overlapping phenomenon. This perspective
dismisses  any connection to a ``sliding friction''. Conversely, other authors
tried to  limit the pedestrians acceleration by introducing ``static friction''
between  the pedestrians and the floor \cite{wang_2019}. This kind of friction,
however,  reduces the effective willings of the pedestrians.  \\

A unique (and universal) set of parameters appears not available yet to our
knowledge.   The reason is that different numerical sets of parameters can lead
to  the same crowd dynamics. Actually, only a small set of dimensionless
``numbers'' control the crowd dynamics \cite{dorso_2019}. These are similar to
those encountered in other active matter systems (P\'eclet number, etc.)
\cite{marchetti_2014}. We may hypothesize that while the dimensionless
parameters provide some kind of control on the  collective behavior in  crowds,
only a few numerical set of parameters can attain an ``individualistic''
meaning.   \\

The numerical parameters for the ``faster is slower'' effect presented by
Helbing and co-workers appears to have some drawbacks
Ref.~\cite{helbing_2000,dorso_2017,dorso_2019}. Although a single  parameter
(say, the desired velocity $v_d$) is numerically varied to  explain the
phenomenon, the researcher may loose sight of the dimensionless  ``numbers''
that truly control the crowd dynamics. The set of parameters in
Ref.~\cite{helbing_2000} also misses the ``faster is faster'' effect reported
to occur at very high pedestrian densities \cite{dorso_2017,haghani_2019}.
Moreover, the empirical fundamental diagram raises as a point of reference  for
the SFM control parameters \cite{helbing_2007,dorso_2017}. \\

The fundamental diagram exhibits the flux behavior for either low density crowds
(with rare contacts between pedestrians) and highly dense crowds (dominated  by
two body interactions). In the second case, crowds experience a flux  slowing
down, but other behaviors are also possible \cite{helbing_2007,lohner_2018}. We
may suspect that the  modeling of the  ``flux slowing down'' within the context
of the SFM will require the proper  setting of the (dimensionless) controlling
parameters. This working hypothesis was already examined in
Ref.~\cite{dorso_2019}, but the parameter exploration was limited to  the
sliding friction, disregarding the body force. In this paper, we widen the
investigation on the set of parameters to include the one  associated to the
body force. We will explore   the complex interplay between  the body force and
the sliding friction among pedestrians. Recall that the  interplay dynamics is
not directly controlled by the set of parameters, but  through dimensionless
``numbers'', where the model parameters appear mixed  between each other. Thus,
this step up offers a challenge to the  ``individualistic'' meaning of the
parameter's set. \\

The paper is organized as follows. We first recall the available  experimental
values on the body force and the sliding friction (see Section
\ref{experimental}). Secondly, we introduce the reduced-in-units SFM and the
three dimensionless numbers that control the crowd dynamics (see Section
\ref{background}). We present our numerical simulations in Section
\ref{results}. For the sake of clarity, this Section is  separated into three
major parts: the bottleneck scenario, the corridor scenario and the comparison
with empirical data in Sections: \ref{bottleneck}, \ref{corridor} and
\ref{Dimensionless}, respectively.  Section \ref{conclusions} opens a detailed
discussion from results in Section \ref{results} and resumes our main
conclusions.

\section{\label{experimental}Experimental background}

The complex behavior of pedestrians features  either his (her) feelings and  the
environmental conditions. The former is expressed, for example, by his  (her)
moving ``attitude'' (say, relaxed, normal, aggressive, etc). The latter brings
out the  observed separation between pedestrians. Also, the ``contacts'' between
individuals produce some kind of ``unwanted'' slowing down. All these  observed
patterns are commonly quantified in the literature into a set of  characteristic
parameters. It is worth mentioning that the empirical data reported in this
Section corresponds to non-panicking situations in which pedestrians  do not
experience high anxiety levels (see below).  The experimental meaning of these
parameters is as follows.

\begin{itemize} \item[(i)] The walking attitude of a pedestrian may appear
somewhat    ``aggressive'' if he (she) reacts actively to unexpected behaviors
\cite{lakoba_2005,helbing_1995}. The smaller the reaction time, the more
aggressive observed posture. The associated parameter to this  behavior is the
relaxation or characteristic time $\tau$  \cite{johansson_2009,helbing_2000}.

\item[(ii)] Despite the reactive attitude $\tau$, the pedestrian adopts a
``free'' (undisturbed) walking speed $v_d$. This speed expresses his (her)
motivation or intention to reach a certain destination (as comfortable as
possible). Observations commonly associate $0.6\,$m/s, $1\,$m/s or $1.5\,$m/s
to relaxed, normal or nervous walking speeds, respectively
\cite{helbing_1995,helbing_2000,li_2015}.  Nevertheless, when pedestrians are in
a panic situation,  they may not reach the ``free'' walking speed, but, $v_d$
may still be   associated to his (her) motivation to get to a target position.

\item[(iii)] The walking speed of pedestrians appears to be lower in a  crowded
walkway with respect to their usually expected ``free'' walking speeds
\cite{weidmann_1992,lakoba_2005}. Pedestrians tend to reduce their speed within
crowded environments because they perceive not enough space for taking a  step
\cite{johansson_2009}. This (perceived) step distance  is therefore an
influential parameter on the pedestrians behavior.  It is known as the
characteristic length $B$.

\item[(iv)] Physical interactions occur in very crowded environments. The
``body force''  and ``sliding friction'' can be introduced straight forward.
This will be done in Section \ref{background}. But it is worth noting that  both
are associated to the moving difficulties (say, slowing down and  obstructions)
observed in contacting pedestrians.

\end{itemize}

Table~\ref{table_data} shows a few empirical values  for the most common
parameters. More data is available throughout the literature (see, for example,
Refs.~\cite{hoogendoorn_2007,seyfried_2007,johansson_2007,moussaid_2009,
luber_2010,seer_2014,li_2015} ). We intentionally omitted data that assumes a
specific mathematical model. The exhibited values should also be considered as a
general purpose approach, since no distinction has been made on age, gender or
cultural habits. \\

\begin{table}
\begin{tabular}{c@{\hspace{6mm}}c@{\hspace{6mm}}c@{\hspace{6mm}}c@{\hspace{6mm}}
c@{\hspace{14mm}}l}
 \hline
 $\tau$[s]   & $m$[kg]     & $v_d$[m/s]  &  $B$[m]  & $k_n$[kg/s$^2$] &  Refs.\\
 \hline
0.61         & ---         & 1.24 & $0.36+1.06\,v$ &  ---                 &  
 \cite{seyfried_2007} \\
0.50$^*$     & 80$^*$      & 1.34 & 0.50           &  ---                 &  
\cite{weidmann_1992,lakoba_2005}\\
---          & $67.5$      & 1.39 &  ---           &  $96.1 + 12694.1\,x$ & 
\cite{song_2019}\\
---          & $67.0$      & 1.39 &  ---           &  $97.0 + 29378.9\,x$ & 
\cite{song_2019}\\

\hline \end{tabular} \caption{The experimental data for the pedestrian
parameters, as explained in  Section~\ref{experimental}. The magnitude $v$ means
the actual pedestrian velocity  (m/s). The magnitude $x$ means the compression
length (m). The upper row for  Ref.~\cite{song_2019} corresponds to data
acquired in winter and the lower row to  data acquired in summer. The asterisk
($^*$) corresponds to reasonable  estimates from the authors. }
\label{table_data}
\end{table}

A first examination of the figures in Table~\ref{table_data} shows that the
choice $\tau\simeq0.6\,$s seems to be a reasonable estimation for the
relaxation time, although this may vary with respect to gender or culture
\cite{siddharth_2018}. Additionally, we confirm that normal pedestrians attain
desired velocities around $1.3\,$m/s. \\

The reports from Refs.~\cite{seyfried_2007,weidmann_1992} do not include any
values for the compressibility $k_n$ since these experiments were carried out
under low density and non-panicking conditions.  The minimum (perceived) step
distance  is $0.36\,$m according to Ref.~\cite{seyfried_2007}, but the
pedestrians seem to require larger  distances when they walk faster. The
commonly accepted value $B\simeq  0.5\,$m is somewhat valid for walking speeds
under $0.5\,$m/s  and non-panicking conditions \cite{seyfried_2007}. Higher
walking speeds (say, $1\,$m/s) will require  a step distance of $1.3\,$m for the
pedestrians to feel that there is enough  space to move along.\\

The reported data from Ref.~\cite{song_2019} correspond to the crowded
environment of the Beijing subway. This environment was not suitable for
providing information on the step distance $B$, but estimates for the  desired
speed and the body compressibility could be achieved. The reported  magnitude
$k_n$ assesses either the clothes and the body compressibility. The  final
value, though, is linearly related to the compression (overlap) $x$. \\

We measured  contact forces during the spring of 2019 at the subway in Buenos
Aires,  Argentina. Our preliminary results show that pedestrians feel
``uncomfortable''  whenever a body force ranging from $5$ to $20\,$N is applied
for at least ten  minutes. Short lasting forces (say, less than 4 minutes) may
also be perceived  as ``uncomfortable'' for values ranging from $10$ to $30\,$N.
We also recorded body  forces up to $60\,$N during very short ``hits''. The
comparison with the  fittings provided by Ref.~\cite{song_2019} shows that these
magnitudes  accomplish densities around 5 people/m$^2$.     \\

The maximum (realistic) overlap may be computed from the Hooke's relation
$F(x)=k_n(x)\,x$ and the compressibility $k_n(x)$ reported in
Table~\ref{table_data}.  An ``uncomfortable'' body force  $10\,$N $-$ $30\,$N
can address overlap values in the  range of $0.030-0.055\,$m. Also, a
``hitting'' force of $60\,$N can address  overlap values between $0.045$ and
$0.065\,$m. Besides, no reliable values for the sliding friction $k_t$ appears
to be available in the literature (to our knowledge).\\

\section{\label{background}Theoretical background}

\subsection{\label{sfm}The Social Force Model}

The Social Force Model (SFM) provides a necessary framework for simulating  the
collective dynamics of pedestrians (\textit{i.e.} self-driven particles). The
pedestrians are considered to follow an equation of motion involving  either
``socio-psychological'' forces and physical forces (say, granular  forces). The
equation of motion for any pedestrian $i$ (of mass $m_i$) reads

\begin{equation}
 m_i\,\displaystyle\frac{d\mathbf{v}_i}{dt}=\mathbf{f}_d^{(i)}+
 \displaystyle\sum_{j=1}^N\mathbf{f}_s^{(ij)}+
 \displaystyle\sum_{j=1}^N\mathbf{f}_g^{(ij)}\label{eqn_motion}
\end{equation}

\noindent where the subscript $j$ corresponds to any neighboring pedestrian or
the walls. The three forces $\mathbf{f}_d$, $\mathbf{f}_s$ and $\mathbf{f}_g$
are different in nature. The desire force $\mathbf{f}_d$ represents the
acceleration (or deceleration) of the pedestrian due to his (her) own will.  The
social force $\mathbf{f}_s$, instead, describes the tendency of the  pedestrians
to stay away from each other. The granular force $\mathbf{f}_g$  stands for both
the sliding friction and the compression between  pedestrians. \\

Notice that these forces are supposed to influence the behavior of the
pedestrians in a similar fashion as mentioned in Section~\ref{experimental}.
Thus, the set of (empirical) parameters described in Section~\ref{experimental}
is expected to be also present in the SFM. These will appear in connection to
the forces. \\

The pedestrians' own will is modeled by the desire force $\mathbf{f}_d$.  This
force stands for the acceleration (deceleration) required to move  at the
desired walking speed $v_d$. This involves, however, a  personal attitude that
makes him (her) appear more or less ``assertive''. As  mentioned in
Section~\ref{experimental}, the reaction time $\tau$ reflects this
attitude. Thus, the desire force is modeled as follows

\begin{equation}
\mathbf{f}_d^{(i)}=m\,\displaystyle\frac{v_d^{(i)}\,
\hat{\mathbf{e}}_d^{(i)}(t)-
 \mathbf{v}^{(i)}(t)}{\tau}
\end{equation}

\noindent where $\hat{\mathbf{e}}(t)$ represents the unit vector pointing to the
target position. $\mathbf{v}(t)$ stands for the pedestrian velocity at time $t$.
\\

The tendency of any individual to preserve his (her) ``private sphere'' is
accomplished by the social force $\mathbf{f}_s$. This force is expected to
prevent the pedestrians from getting too close to each other (or to the walls)
in a any environment. The model for this kind of  ``socio-psychological''
behavior is as follows

\begin{equation}
 \mathbf{f}_s^{(i)}=A\,e^{(R_{ij}-r_{ij})/B}\,\hat{\mathbf{n}}_{ij}
 \label{eqn_social}
\end{equation}

\noindent where $r_{ij}$ means the distance between the center of mass of the
pedestrians $i$ and $j$, and $R_{ij}=R_i+R_j$ is the sum of the pedestrians
radius. The unit vector $\hat{\mathbf{n}}_{ij}$ points from pedestrian $j$ to
pedestrian $i$, meaning a repulsive interaction.\\

The net distance (overlap) $|R_{ij}-r_{ij}|$ scales to the parameter $B$ in the
expression (\ref{eqn_social}). This parameter plays the role of a fall-off
length within the model, and thus, it may be somewhat connected to the
(perceived) step distance mentioned in Section~\ref{experimental}.  Besides, the
parameter $A$ reflects the intensity of the social repulsion. \\

The granular force (say, the sliding friction plus the body force) reflects  the
moving difficulties encountered in very crowded environments. The  expression
for the granular force has been borrowed from the granular  matter field, the
mathematical expression reads as follows

\begin{equation}
 \mathbf{f}_g^{(ij)}=k_t\,g(R_{ij}-r_{ij})\,
(\Delta\mathbf{v}^{(ij)}\cdot\hat{\mathbf{t}}_{ij})\,\hat{\mathbf{t}}_{ij}+
k_n\,g(R_{ij}-r_{ij})\,
\,\hat{\mathbf{n}}_{ij}\label{eqn_friction}
\end{equation}

\noindent where $g(R_{ij}-r_{ij})$ equals $R_{ij}-r_{ij}$ if $R_{ij}>r_{ij}$ and
vanished otherwise. $\Delta\mathbf{v}^{(ij)}\cdot\hat{\mathbf{t}}_{ij}$
represents the relative tangential velocities of the sliding  bodies (or between
the individual and the walls).    \\

The sliding friction occurs in the tangential direction while the body force
occurs in the normal direction. Both are assumed to be linear with respect to
the net distance between contacting pedestrians. The sliding friction is also
linearly related to the difference between the (tangential) velocities. The
coefficients $k_t$ (for the sliding friction) and $k_n$ (for the  body force)
are supposed to be related to the areas of contact and the clothes  material,
among others. \\

We stress that the expression (\ref{eqn_friction}) assumes fixed values for
$k_t$ (and $k_n$). This may not be completely true according to
Table~\ref{table_data}. The local density (and thus, the pedestrians'
compression) may affect the compressibility parameter $k_n$ by more than an
order  of magnitude. We will vary $k_n$ (and $k_t$) in order to explore this
phenomenon. \\

\subsection{\label{parameters}The set of parameters}

The numerical set of parameters may affect the dynamics of the  pedestrians.
Some values, however, yield similar collective dynamics. We introduce
dimensionless magnitudes  and proceed straightforward as indicated in
\ref{appendix1}. We realize from  \ref{appendix1} that only three
(dimensionless) parameters  are the true  ``control'' parameters for the
collective dynamics. These are $\mathcal{A}$,  $\mathcal{K}$, $\mathcal{K}_c$ as
defined in \ref{appendix1}. Recall that  $\mathcal{A}$ and $\mathcal{K}$ are
precisely the same as in  Ref.~\cite{dorso_2019}, but a novel $\mathcal{K}_c$
parameter has been  introduced due to the body force.   \\

\begin{figure}[!htbp]
\centering
 
\subfloat[\label{venn_2parameters}]
{\includegraphics[width=0.49\columnwidth]{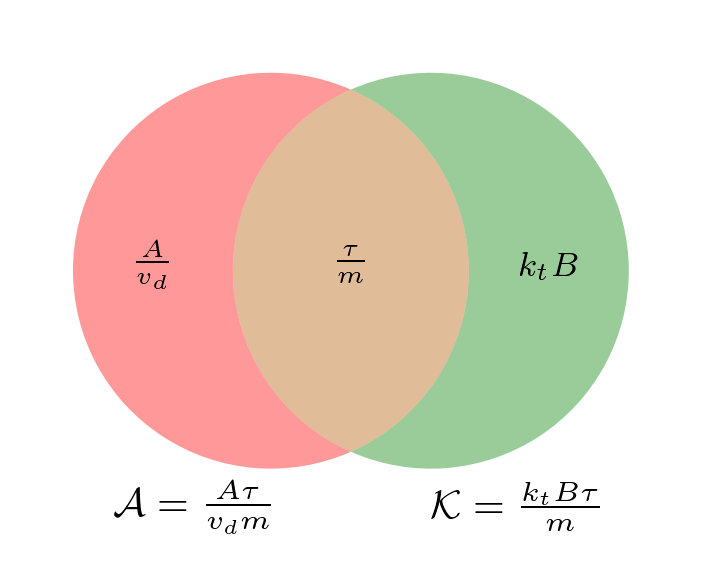}} \hfill
\subfloat[\label{venn_3parameters}]
{\includegraphics[width=0.49\columnwidth]{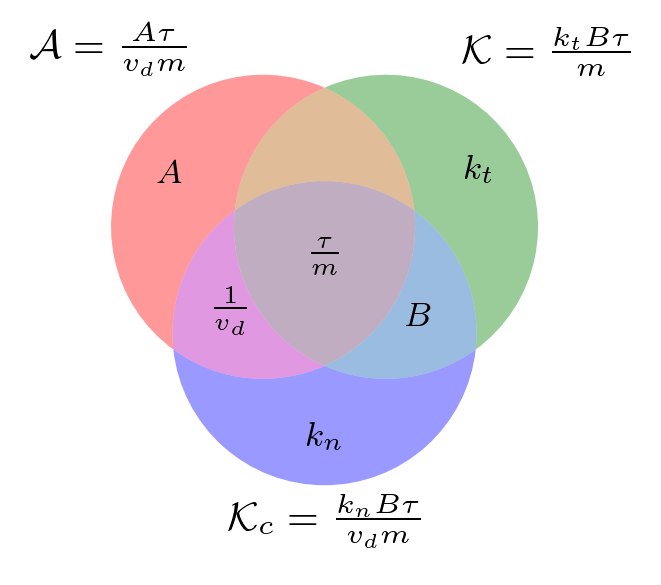} }\\
\caption[width=0.47\columnwidth]{Venn diagrams for the dimensionless parameters
appearing in the equation of motion (\ref{eqn_motion}) (see
Appendix~\ref{appendix1} for details). The sets correspond to
$\mathcal{A}=\{\tau/m,1/v_d,A\}$, $\mathcal{K}=\{\tau/m,B,k_t\}$ and
$\mathcal{K}_c=\{\tau/m,1/v_d,B,k_n\}$. (a) The Venn diagram representation if
no  body force is introduced in the SFM (only sets $\mathcal{A}$ and
$\mathcal{K}$,  as in Ref.~\cite{dorso_2019}). (b) The Venn diagram
representation for the sets  $\mathcal{A}$, $\mathcal{K}$ and $\mathcal{K}_c$.}
\label{venn_diagram} 
\end{figure}

The logical relations between the dimensionless parameters and the
``individual''  parameters can be studied by means of the Venn diagrams
exhibited in  Fig.~\ref{venn_diagram}. The parameters $\mathcal{A}$,
$\mathcal{K}$ and  $\mathcal{K}_c$ are represented as intersecting sets, and the
``individual''  parameters are represented as elements within each set. The
shared elements  between $\mathcal{A}$, $\mathcal{K}$ and $\mathcal{K}_c$ are
placed inside the  intersecting regions. \\

A first inspection of the diagrams in Fig.~\ref{venn_diagram} shows that the
relaxation time (per unit mass) $\tau/m$ is always a common parameter to all
sets, regardless of the body force. This means that the ``assertive'' attitude
of the pedestrian, addressed by the reaction time (see
Section~\ref{experimental}), applies to all stimuli and the own willings. The
role of $\tau$ has already been discussed in
Refs.~\cite{johansson_2009,dorso_2019}.    \\

Fig.~\ref{venn_2parameters} represents the situation when $\mathcal{K}_c$ is
absent. Notice that $A/v_d$ or $k_t B$ may control the collective  dynamic, in
spite of the ``assertive'' attitude. The individual character of  $v_d$ or $B$
appears somehow ``loosely'' in the crowd dynamic. We mean by  ``loosely'' that
any numerical set for these parameters may be  counterbalanced by the right
choice of $A$ or $k_t$, keeping the  collective dynamic (qualitatively)
unchanged.  \\

Fig.~\ref{venn_3parameters} provides a picture of the parameters' relations
after introducing $\mathcal{K}_c$. Surprisingly, $\mathcal{K}_c$ appears as a
wider set (say, a four elements set) than $\mathcal{A}$ or $\mathcal{K}$ (three
elements' sets). It shares the parameter $v_d$ with $\mathcal{A}$ and the
parameter $B$ with $\mathcal{K}$. The practical consequence to these (logical)
relations is that $v_d$ or $B$ affect simultaneously two ``control'' parameters
of the collective dynamics. Conversely, either $v_d$ and $B$ may counterbalance
$\mathcal{K}_c$ in order to keep the collective dynamic (qualitatively)
unchanged.  \\

We confirm from these diagrams that no univocal relations can be established
between the individual parameters and the collective dynamics (in a crowded
environment). The presence of the body force moves the dynamics to a more
complex context. We will investigate this context in Section~\ref{results}. \\

\subsection{\label{blocking_clusters} Blocking clusters}

A characteristic feature of pedestrian dynamics is the formation of clusters.
Clusters of pedestrians can  be defined as the set of individuals that for any
member of the group (say, $i$) there exists at least another member belonging to
the same group ($j$) in contact with the former.  Thus, we define a ``granular
cluster'' ($C_g$) following the mathematical formula given in
Ref.~\cite{strachan1997fragment}

\begin{equation}
C_g:P_i~\epsilon~ C_g \Leftrightarrow \exists~ j~\epsilon~C_g / r_{ij} 
< (R_i+R_j) \label{ec-cluster}
\end{equation}

where ($P_i$) indicate the \textit{ith} pedestrian and $R_i$ is his (her) radius
(shoulder width). That means, $C_g$ is a set of pedestrians that interact not
only with the social force, but also with physical forces (\textit{i.e.}
friction force and body force). A ``blocking cluster'' is defined as the subset
of clusterized particles (granular cluster) closest to the door whose first  and
last component particles are in contact with the walls at both sides of the door
~\cite{dorso_2005}. This clogging structure is responsible for worsening the
evacuation performance.

\section{\label{hypotheses}Working hypotheses and procedures}

\subsection{Working hypotheses}

Our working hypotheses are similar to those mentioned in
Ref.~\cite{dorso_2019},  although the presence of the body force opens inquiries
about the proper  exploration of the parameter space. To be precise

\begin{itemize}  \item[(a)] The ``faster-is-slower'' (or the ``faster-is-
faster'') phenomenon  occurs when varying the values of $v_d$ (see
Section~\ref{bottleneck}). This is  equivalent to vary simultaneously
$\mathcal{A}$ and $\mathcal{K}_c$ by the same  amount in the (dimensionless)
parameter space (see Section~\ref{parameters}). We  may visualize this sampling
procedure as moving along a straight line in the  space
$(\mathcal{A},\mathcal{K},\mathcal{K}_c)$. Further variations of  $k_n$ turns
the sample points out of this line, but on the plane of constant
$\mathcal{K}_c$. We will step up $k_n$ by several orders of magnitude in order
to get the big picture of this surface.

\item[(b)] We will consider the parameter set from Ref.~\cite{helbing_2000}  as
a starting point for exploring the parameter space. The reason for this is  that
a complete set of experimental parameters is still not available (to our
knowledge). We are aware, though, of the drawbacks of this choice, say, the
unrealistic meanings for the length $B$ and the compressibility $k_n$ (see
Section~\ref{experimental}). But these are irrelevant in the context of the
(dimensionless) parameter space and the corresponding collective dynamics (see
Section~\ref{introduction}).

\item[(c)] We will explore crowd densities allowing body compressions (overlaps)
up to $0.1\,$m. This corresponds to remarkably high densities, since our own
(real- life) estimates do not surpass $0.065\,$m (see
Section~\ref{experimental}). Therefore, the extremely high density  scenarios
(say, above $0.065\,$m) will be considered for the purpose of a  tendency, but
not as a common real-life situation. No casualties due to high  pressures will
be further considered.

\end{itemize}

We stress the fact that our concern is placed on the collective dynamics of
crowded environments. We will work on the hypothesis that any ``reasonable''
parameter set should reproduce the collective behavior (say, the slowing down in
the fundamental diagram; see Section~\ref{Dimensionless}). We will not attempt
to optimize parameters from other objective functions. \\

We will further sustain the hypothesis of soft matter in the context of crowded
environments. The body compression and the sliding friction parameters are
supposed to be connected in some way under this hypothesis. However, we will not
introduce any direct link between $\mathcal{K}$ and $\mathcal{K}_c$. We will
investigate the interplay between both in Section~\ref{Dimensionless}.\\

In order to keep the model as simple as possible, we will only consider
isotropic pedestrian interactions, since anisotropic interactions do not play a
relevant role at high densities. \\

\subsection{Procedures} 

The SFM was implemented on the LAMMPS simulation software \cite{plimpton}.
Additional modules for LAMMPS were also written in C++ in order to expand the
software capabilities. All these were able to run in a high performance
environment (HPC). \\

The implemented SFM parameters were the same as those in
Ref.~\cite{helbing_2000} (at the beginning of the exploratory procedure only).
But the pedestrian's mass and radius were set to the more realistic values of
$70\,$kg and $0.23\,$m, respectively. The force interactions between
pedestrians were limited, however, to a cut-off distance of $0.88\,$m for
attaining a \textit{privacy sphere} that excludes second neighbors. The desired
velcity was always set to $1\,$m/s in the corridor situation. Besides, the
explored values of $v_d$ for the bottleneck scenario ranged from $1\,$m/s to the
extremely anxious situation of $10\,$m/s. \\

The Eq.~(\ref{eqn_motion}) was numerically integrated by means of the velocity
Verlet algorithm, with a timestep of $10^{-4}$ seconds. The pedestrians
positions and velocities were recorded every $0.05\,$sec, but post-processing
computing was done over samples acquired at least every $2\,$sec., in order to
avoid  data correlations. Those pedestrians leaving the simulations box were
re-introduced into the box, on the opposite side (periodic boundary
conditions). We only omitted this mechanism when computing the evacuation time
for the bottleneck geometry. \\

The post-processing computing was assisted by Python functions. The NetworkX
package was used among others.   \\

We warn the reader that, for simplicity, we will not include the units
corresponding to the numerical results. Remember that the friction coefficient
has units $\left [ k_t \right]=$Kg~m$^{-1}$~s$^{-1}$, 
the body stiffness coefficient $\left [ k_n \right]=$Kg~s$^{-2}$, 
the density $\left [ \rho \right]=$p~m$^{-2}$ and the flow $\left [ J \right
]=$p~m$^{-1}$~s$^{-1}$.\\

\section{\label{results}Results}

\subsection{\label{bottleneck} Bottleneck}

We present in this section the results corresponding to the bottleneck geometry.
We show the consequences of modifying the body force coefficient $k_n$ on the
evacuation dynamics. Recall that this coefficient is associated  to the
compression of the human body. \\

\begin{figure}[htbp!]
\centering
\includegraphics[width=0.7\columnwidth]{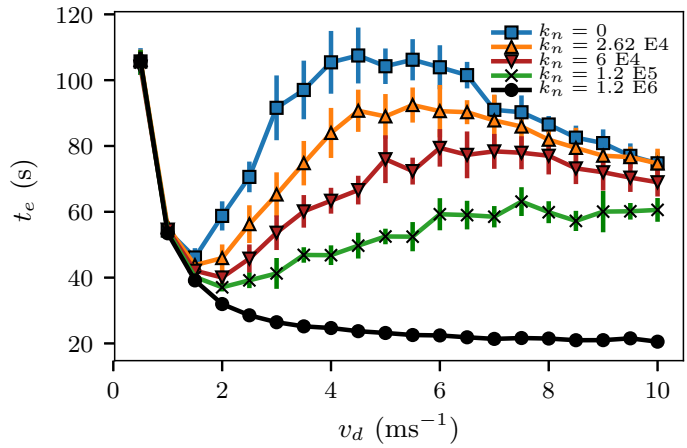}

\caption{\label{vd_vs_te}Mean evacuation time (s) vs. the pedestrian’s desired
velocity (m/s) for a bottleneck. The room was 20 m x 20 m size. The door was
0.92 m width (two pedestrians' width). Mean values were computed from 10
evacuation processes. 225  pedestrians were initially placed in a square lattice
with a random initial  velocity. Each process was finished when 158 pedestrians
left the room. The  different symbols indicate the $k_n$ value corresponding to
the body force (see  the label). The crosses correspond to the Helbing's
original SFM parameter, the  up-triangles correspond to the value measured in
Ref.~\cite{melvin1988aatd}, squares correspond  to zero body force and circles
correspond to an extreme value of stiffness (one  order of magnitude higher than
the original SFM). The   down triangles correspond to an intermediate value
between the empirical value  presented in Ref.~\cite{melvin1988aatd} and the one
provided by Helbing in  Ref.~\cite{helbing_2000} }

\end{figure}

Fig.~\ref{vd_vs_te} shows the evacuation time as a function of the pedestrian's
desired velocity for different values of $k_n$. The evacuation  time is defined
as the time lapsed until 80\% of the pedestrians have  left the room. In this
section we will focus on the evacuation time for  $2\,$m/s$<v_d<10\,$m/s.\\

Three behavioral patterns can be distinguished in Fig.~\ref{vd_vs_te}.  Each
pattern can display a positive slope, a negative slope or both. The interval in
which the slope is positive means that the  harder the pedestrians try to get
out (higher $v_d$), the  longer it takes them to evacuate. This is the Faster-
is-slower  (FIS) regime. Conversely, the interval in which the slope is
negative  corresponds to a Faster-is-Faster (FIF) regime (the  harder they try,
the quicker they leave).\\

Fig.~\ref{vd_vs_te}  shows either FIS or FIF, and a FIS+FIF pattern for desired
velocities $v_d>1.2\,$m/s. The  evacuation time attains a FIS+FIF pattern for
compression coefficients below  $k_n=1.2\,$E5. This means that ``soft''
individuals can attain this  behavioral pattern.  Notice that higher values of
$k_n$ allow  only FIS or FIF patterns. For the highest explored value
$k_n=1.2\,$E6, no  FIS can be seen at all. Besides, the evacuation pattern for
$k_n = 0$ and $k_n  = 2.6\,$E4 are very similar since the body force intensity
is of the same order or less than the social  force for these stiffness values.
\\

\begin{figure}[!htbp]
\centering
\subfloat[]{\includegraphics[width=0.7\columnwidth]{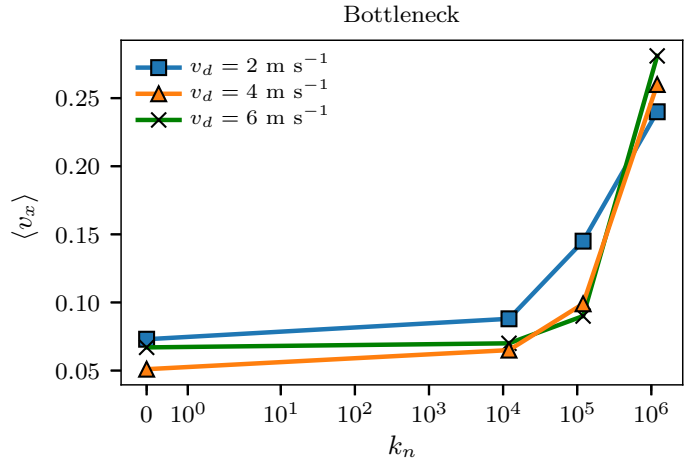}}\\
\caption[width=0.47\columnwidth]{Mean velocity in the $x-$direction as a
function  of the stiffness level $k_n$ for different desired velocities (see
label). The  data corresponds to a bottleneck with periodic boundary conditions
(re-injecting  pedestrians). The average was taken every five seconds once the
crowd reaches the  stationary state ($t=20\,$s) until the end of the simulation
($t=1000\,$s).  Color online only. }
\label{kn_vs_vx_bottleneck}
\end{figure}

Despite the presence of the FIS or FIF pattern, the evacuation time at a  fixed
value of $v_d$ decreases for increasing values of $k_n$ (within the examined
interval).  This means that stiffer pedestrians evacuate faster than soft
pedestrians.  To further investigate this  phenomenon, we computed the mean
velocity of the whole crowd as a function of the  stiffness $k_n$.
Fig.~\ref{kn_vs_vx_bottleneck} shows the actual mean velocity in the
$x-$direction as a function of the stiffness $k_n$ for three different desired
velocities. In all cases, the velocity increases with the stiffness. The most
significative  increment of velocity is in the interval $k_n>\,10^{4}$. \\

The existence of the FIF phenomenon for only ``stiff'' individuals opens many
questions  on the microscopic dynamics of pedestrians. We may presume that
contacts  between pedestrians are quite different for soft individuals than for
stiff  individuals. Thus, we proceed to study the dynamics of contacting
pedestrians, regardless of the overlap  effects. We will assimilate the
pedestrians as nodes and the whole crowd as a  network. We will link any two
individuals whenever they get in physical  contact (\textit{i.e.} $r_{ij} \leq
R_{ij}$).\\

\begin{figure}[!htbp]
\centering
\subfloat[]{\includegraphics[width=0.49\columnwidth]
{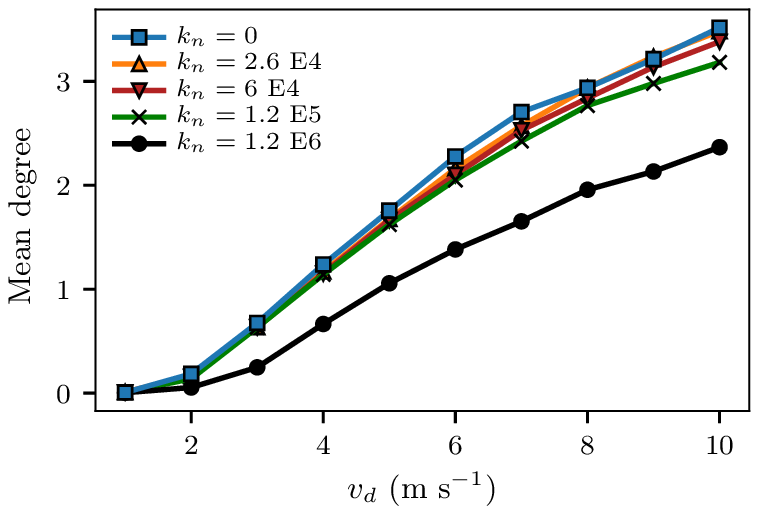}\label{degree_vd}}\ 
\subfloat[]{\includegraphics[width=0.49\columnwidth]
{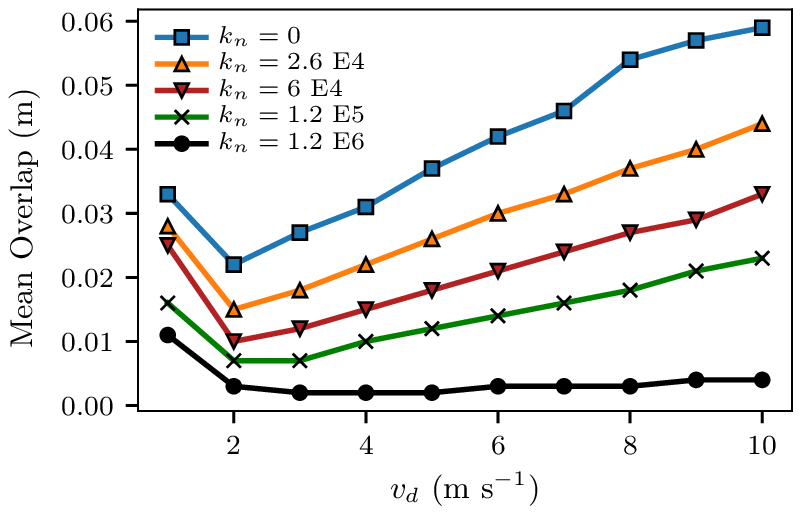}\label{overlap_vd}}\\
\caption[width=0.47\columnwidth]{(a) Mean degree as a function of the
pedestrian’s desired velocity. (b) Mean overlap as a function of the
pedestrians desired velocity. Each symbol indicates the $k_n$ value
corresponding to the body force (see the labels). The data corresponds to a
bottleneck with periodic boundary conditions (re-injecting pedestrians). The
average was taken over time and the pedestrians in the bottleneck. The sampling
was done every five seconds once the crowd reached the stationary state (say,
$t=20\,$s) until the end of the simulation ($t=1000\,$s). Color online only.}
\label{degree_overlap_vd}
\end{figure}

Fig.~\ref{degree_vd} shows the mean degree of the contact network as a function
of the desired velocity. The degree of a node is defined as the number of links
that connects this node to any other node. This means, the number of pedestrians
that are in physical contact with a given pedestrian. The mean degree is the
average of the degree over all the nodes (pedestrians) and over the whole
sampled interval. We computed mean values only after the system reached the
stationary  state, that is, after a well-formed bulk has been established.\\

Notice that the mean degree increases as $v_d$ increases, as expected. This
expresses the fact that higher $v_d$ values accomplish higher densities, forcing
individuals to touch each  other. For a given $v_d$, the mean degree reduces as
the $k_n$  value increases. A noticeable decrease in the mean degree can be seen
for the  highest explored value of $k_n$. This opens the question on how would
this affect the sliding friction among pedestrians.\\

A more detailed insight into the contact dynamics can be acquired from
Fig.~\ref{overlap_vd}. The overlap between individuals is shown as a function of
$v_d$. Recall from Section \ref{sfm} that the overlap is defined as
$R_{ij}-r_{ij}$ where $R_{ij}$, is the sum of radius of particle $i$ and
particle $j$ and $r_{ij}$  is the distance between both particles. Except for
very low desired velocities (say, $v_d<2\,$m/s), we can see that the mean
overlap is an increasing function of $v_d$ (withing the studied range of $v_d$
and $k_n$).\\

The regime for $v_d<2\,$m/s is qualitatively different from the clogged regime
since most of the  pedestrians are not touching each other (as was already
observed in Ref~\cite{dorso_2005,dorso_2011}). The only pedestrians who touch
each  other are the ones who are being re-injected on the opposite side of the
room. These collide with the bulk, providing some kind of overlap during very
short time. We will not analyze this regime.\\

Besides, for a given $v_d>2\,$m/s, the overlap increases as the $k_n$ value
decreases.  This can be explained by considering the bulk at a (quasi)
equilibrium situation.  The social and compression forces counterbalance the
desired force. Thus, for any fixed $v_d$, the product $k_n \times
(R_{ij}-r_{ij})$ remains (almost) fixed. Any decrease in $k_n$ allows a more
significant intrusion. This (partially) supports the argument that the sliding
friction should weaken for stiffer pedestrians (say increasing $k_n$ values).\\

The sliding friction reduction appears as the first feature for enhancing the
overall evacuation performance. Either reducing the mean overlap and the mean
degree tend to diminish the mean sliding  friction within the crowd.  Notice,
however, that switching from a FIS regime (positive slope) to a  FIF  regime
(negative slope) in Fig.~\ref{vd_vs_te} appears as a more complex  phenomenon.
We will focus on this issue in an upcoming investigation.\\

\begin{figure}[!htbp]
\centering
\subfloat[]{\includegraphics[width=0.49\columnwidth]
{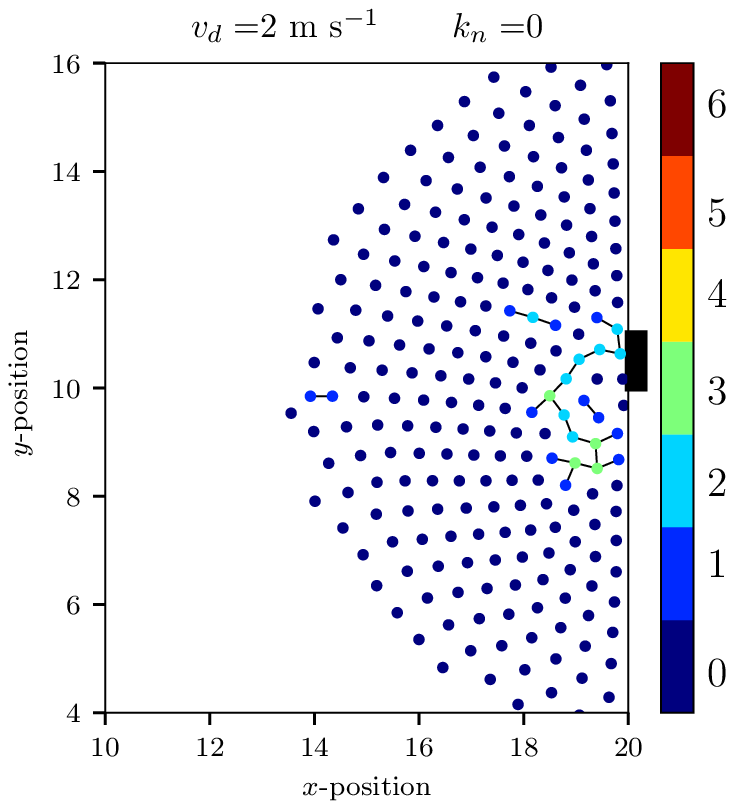}\label{network_vd2_kn0}}\ 
\subfloat[]{\includegraphics[width=0.49\columnwidth]
{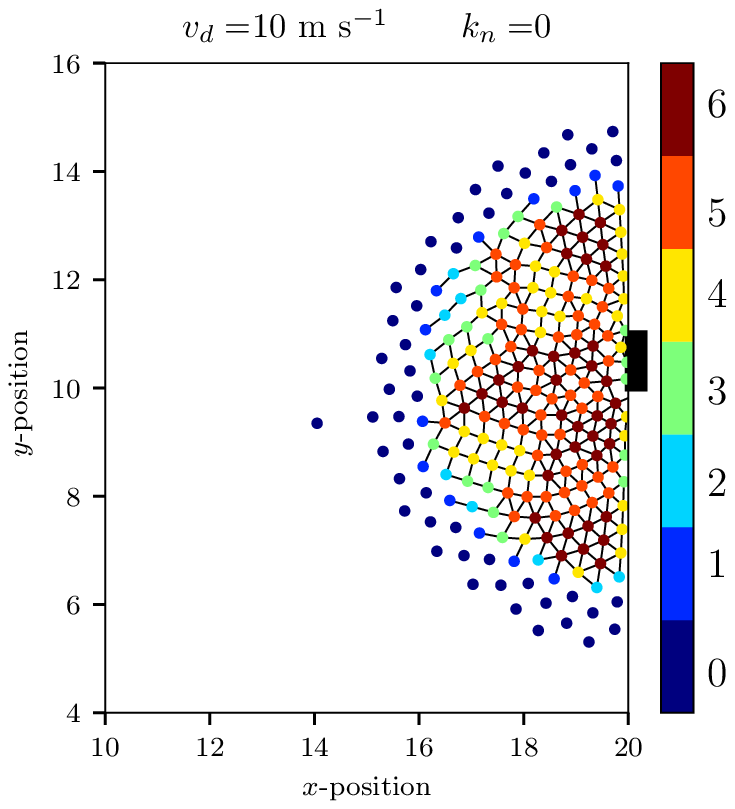}\label{network_vd10_kn0}}\\
\subfloat[]{\includegraphics[width=0.49\columnwidth]
{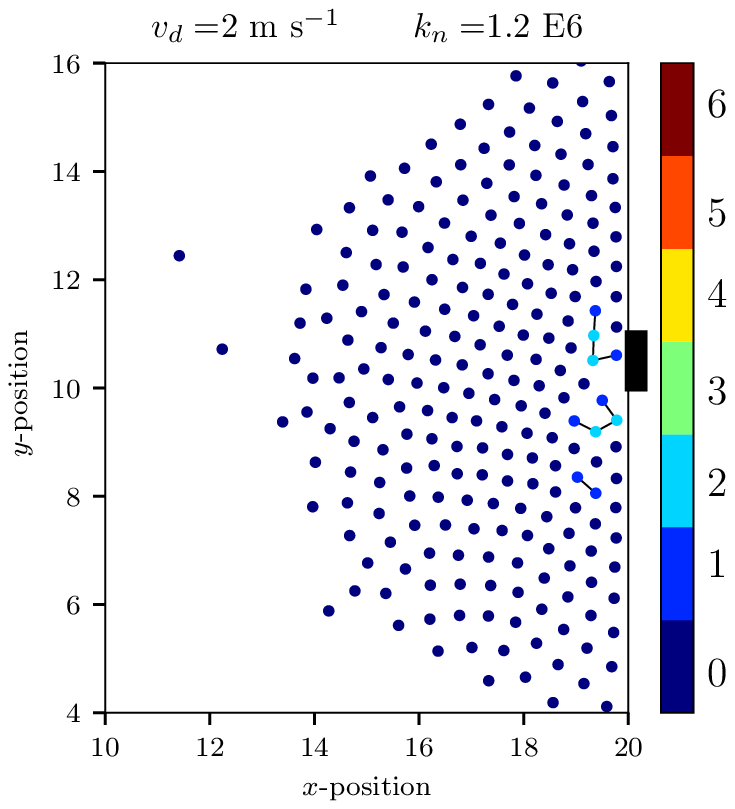}\label{network_vd2_kn1200000}}\ 
\subfloat[]{\includegraphics[width=0.49\columnwidth]
{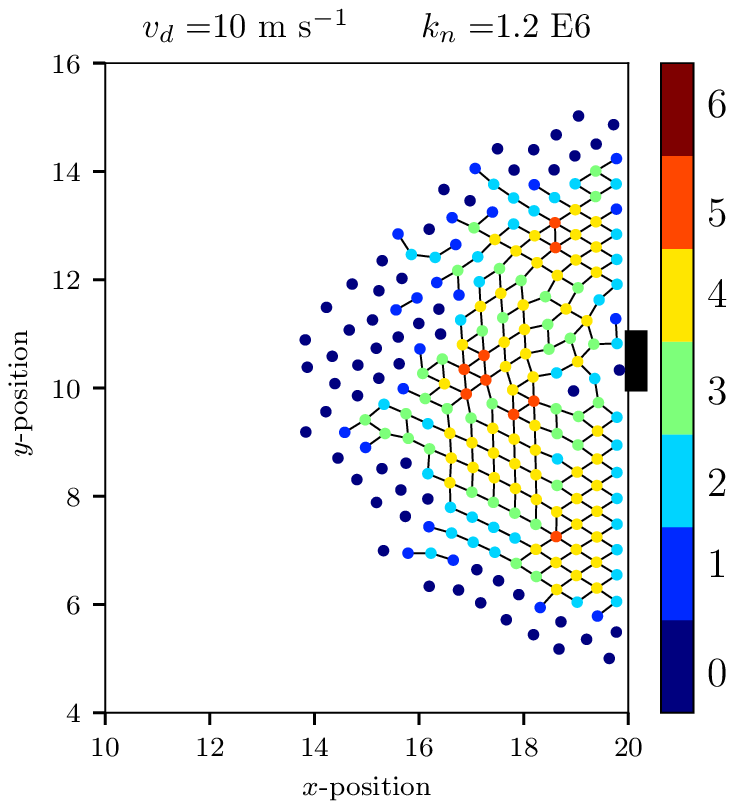}\label{network_vd10_kn1200000}}\\

\caption[width=0.47\columnwidth]{Snapshots of the contact networks of a 225
pedestrian evacuation through a bottleneck. The door is placed at
$(x,y)=(20,10)\,$m, the width of the door is 0.92$\,$m (equivalent to 2
pedestrian\textsc{\char13}s diameter). The lines that connect the nodes
(pedestrians) represent the contact between them. The color represents the
degree (the number of pedestrians with which it is connected). (a) and (b)
correspond to a simulation without body force with $v_d=$2 and $v_d=$10
respectively. (c) and (d) correspond to simulations with $k_n=1.2\,$E6 with
$v_d=$2 and $v_d=$10 respectively. The black rectangle at the right
represents the exit door. Color online only.}
\label{network_bottleneck}
 \end{figure}

The body force  has a notorious impact in the number of pedestrians touching
each other (say,  the degree). This is clearly depicted in
Fig.~\ref{network_bottleneck} where  four different configurations of the
evacuation dynamics are shown. The  configurations represent 225 pedestrians
trying to escape through a door (see  caption for details). The colors
correspond to the degree of each node  (pedestrian), and the lines between
pedestrians represent the contacts among  them.\\

The four configurations corresponds to two different $v_d$ and two different
$k_n$ values (say, the  minimum and maximum explored values).
Fig.~\ref{network_vd2_kn0} and Fig.~\ref{network_vd10_kn0} show snapshots for
$k_n=0$, at the desired velocities of 2$\,$m/s and 10$\,$m/s, respectively.
Fig.~\ref{network_vd2_kn1200000} and Fig.~\ref{network_vd10_kn1200000} show
similar situations, but for  $k_n=1.2\,$E6. As expected, increasing the desired
velocity compresses the crowd towards the exit. \\

The four snapshots in Fig.~\ref{network_bottleneck} confirm (visually) the fact
that more rigid pedestrians ease the crowded environment, widening the occupied
region. At $v_d=10\,$m/s (the maximum explored velocity), it can hardly be found
pedestrians with degree 6 when $k_n=1.2\,$E6, while a lot of them are present
for $k_n$=0.\\

\begin{figure}[htbp!]
\centering
\includegraphics[width=0.7\columnwidth]
{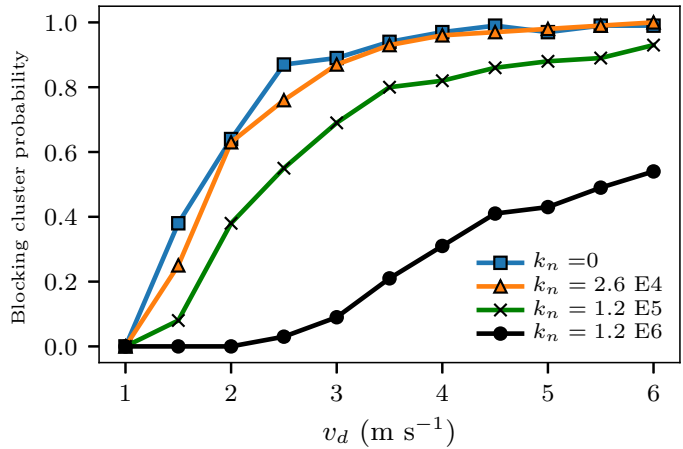}
\caption{\label{pbc_vs_vd_multi_kn} Blocking cluster probability as a function 
of $v_d$ for different stiffness levels (see label). The probability is 
calculated as the amount of time a blocking cluster is present divided by the 
overall simulation time. The situation corresponds to a bottleneck with 225 
pedestrians under periodic boundary conditions (re-injection of pedestrians once
 they left the room). The sampled interval was set to $t_f =$ 1000$\,$s. 
 Color online only. }
\end{figure}

The clusterization of the pedestrians has a significant impact on the blocking
clusters (the group of pedestrians that clog the exit).
Fig.~\ref{pbc_vs_vd_multi_kn} shows the blocking cluster probability as a
function of the desired velocity for different $k_n$ values (see caption for
details). The blocking clusters become more probable for high desired velocity,
since the clogged area gets more compact as $v_d$ increases (for any fixed value
of $k_n$). But, the most remarkable fact in Fig.~\ref{pbc_vs_vd_multi_kn} is
that increasing the body stiffness reduces the blocking cluster probability for
any fixed value of $v_d$. Recall from Ref.~\cite{dorso_2005} that the evacuation
time is controlled by the blocking. Thus, increasing the body stiffness affects
the presence of the blocking clusters, and consequently improves the evacuation
time. This raises as a second feature for enhancing the overall evacuation
performance. \\

We may summarize this Section as follows. We explored the parameter space along
$v_d$ and $k_n$ for the bottleneck situation. This is similar to explore the
dimensionless plane $(\mathcal{A},\mathcal{K}_c)$. Soft pedestrians attain a FIS
or FIS+FIF evacuation pattern but, stiff pedestrians (for any fixed
$\mathcal{A}$ value within the explored range) exhibit a single FIF pattern.
Stiffness affects either the presence of the blocking clusters and the
pedestrians overlap. The less they overlap, the less intense becomes the sliding
friction among them. Additionally, the fewer the blocking clusters the easier
they get out. \\

\subsection{\label{corridor} Corridor}

We present in this section the results corresponding to the corridor geometry.
We show the effects of varying the body force coefficient $k_n$ on the
collective dynamics. We will keep $v_d$ (or $\mathcal{A}$)  fixed along this
Section. \\

\begin{figure}[!htbp]
\centering
    
\subfloat[]{\includegraphics[width=0.49\columnwidth]
{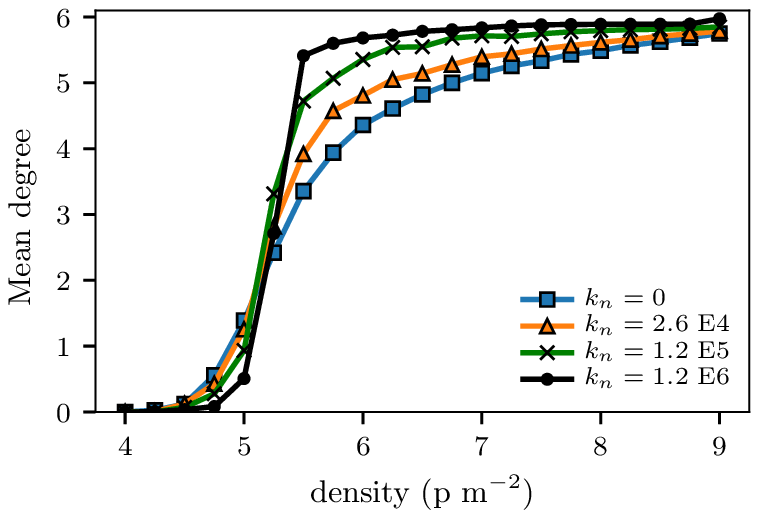}\label{degree_dens}}\ 
\subfloat[]{\includegraphics[width=0.49\columnwidth]
{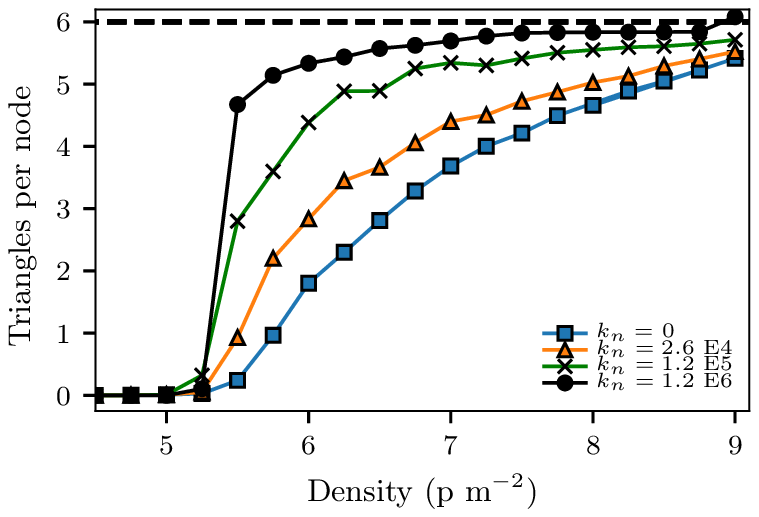}\label{triangles}}\\
\caption[width=0.47\columnwidth]{(a) Mean degree as a function 
of the global density for different $k_n$ values. (b)  Triangles per node as a 
function of the global density. The global density is the total number of 
pedestrians per unit area. The mean values are averages over all the pedestrians 
and over time once the system reached the stationary state. The measurements 
correspond to a corridor of 28~m $\times$ 22~m with periodic boundary conditions 
and $v_d=1$. Color online only.}
\label{network_corridor}
\end{figure}

We first computed the contact network in the same way  as in the bottleneck
geometry. Fig.~\ref{degree_dens} shows the  mean degree as a function of the
global density for different $k_n$ values (see  caption for details). The mean
degree vanishes at very low  densities because the pedestrians do not touch each
other. When the density  surpasses 4.5, a few pedestrians start to touch each
other,  raising the mean degree. As the density continuous increasing,  the mean
degree approaches the asymptotic value  of six. Degree six corresponds to the
maximum packing density for identical hard disks.\\

It is worth noting from Fig.~\ref{degree_dens} that ``stiff'' pedestrians
exhibit a sharper transition to the maximum packing density than  the ``soft''
pedestrians. We further checked this phenomenon by counting the  number of loops
of three nodes (say, the triangles) present in the network.  This magnitude was
reported to be a sensitive magnitude for characterizing the  jamming transition
in a compressed granular packing (see  Ref.~\cite{pugnaloni_2013} for details).
As can be seen in  Fig.~\ref{triangles}, the ``stiff'' pedestrians share more
triangles  than the ``soft'' ones.\\

\begin{figure}[!htbp]
\centering
\subfloat[]{\includegraphics[width=0.49\columnwidth]
{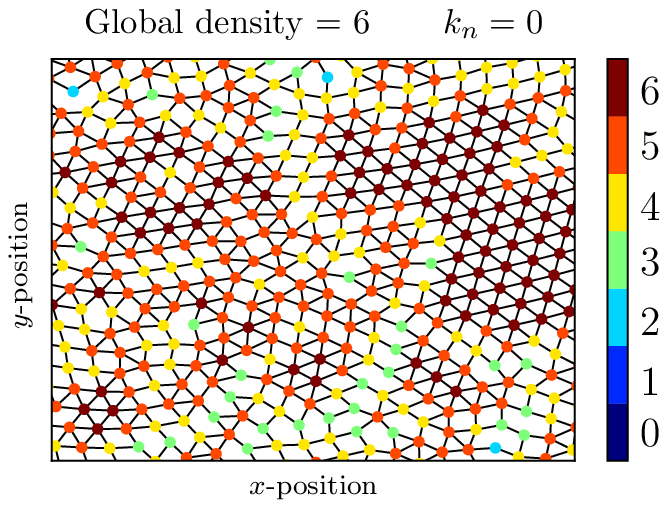}\label{network_d6_kn0}}\ 
\subfloat[]{\includegraphics[width=0.49\columnwidth]
{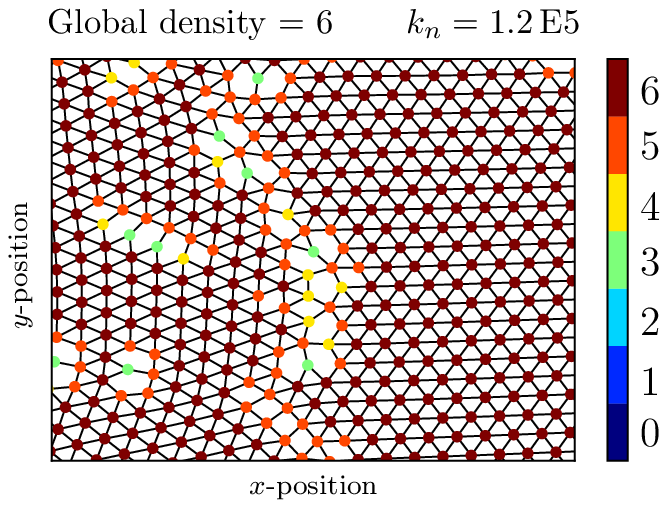}\label{network_d6_knE5}}\\
\caption[width=0.47\columnwidth]{Contact network of the pedestrians along the 
corridor at time $t=50\,$s. The global density was $\rho=6$.  The lines that 
connect the nodes (pedestrians) represent the contacts between them. The colors 
stand for the degree of the node (the number of pedestrians that are in contact 
with him/her). The corridor was 28~m $\times$ 22~m with periodic boundary 
conditions and $v_d=1$. (a) corresponds to a simulation without body force and 
(b) corresponds to a simulation with $k_n=1.2$~E5. The friction coefficient and 
the other SFM parameters are the same as in Section \ref{bottleneck}. Color 
online only.}
\label{network_corridor}
\end{figure}

In order to get a better insight on how the pedestrians contact to each other,
we present in Fig.~\ref{network_corridor} two snapshots of the corridor at the
stationary situation. Fig.~\ref{network_d6_kn0} corresponds to  $k_n=0$, while
Fig.~\ref{network_d6_knE5} corresponds to $k_n=1.2$~E5. The  former shows a
somewhat disordered network, while the latter exhibits an almost  completely
ordered lattice. The missing triangles in Fig.~\ref{network_d6_kn0}  are
replaced by other polygons of more than three edges. \\

We find these topological magnitudes useful for comparing the  pedestrian
behavior in the  corridor geometry with respect to the bottleneck geometry. A
re-examination of  Fig.~\ref{degree_dens} (corridor) and Fig.~\ref{degree_vd}
(bottleneck) reveal  that the pedestrian stiffness $k_n$ affects differently the
way they contact  each other. The mean degree increases for ``stiff''
pedestrians moving along  the corridor as the density increased with a sharp
increase at  $5\,$p/m$^2$. Conversely, the mean degree  increases for ``soft''
pedestrians in the bottleneck situation.\\

Fig.~\ref{network_bottleneck} and  Fig.~\ref{network_corridor} illustrate the
connectivity differences between  the bottleneck and the corridor situation. The
bulk in  Fig.~\ref{network_bottleneck} appears more heavily connected among
``soft''  pedestrians than among ``stiff'' pedestrians (see both snapshots at
$v_d=10$).  The opposite occurs in Fig.~\ref{network_corridor}. This discrepancy
seems to  be related to the boundary conditions, since the same SFM parameters
were  applied on both situations. We may speculate that this phenomenon occurs
because, in the corridor, the lateral walls act like a  confining barrier that
forces the ``stiff'' pedestrian to increase his (her) contacts. On the contrary,
no real confining walls exist in the bottleneck situation (regardless the side
walls). \\

\begin{figure}[htbp!]
\centering
\includegraphics[width=0.7\columnwidth]
{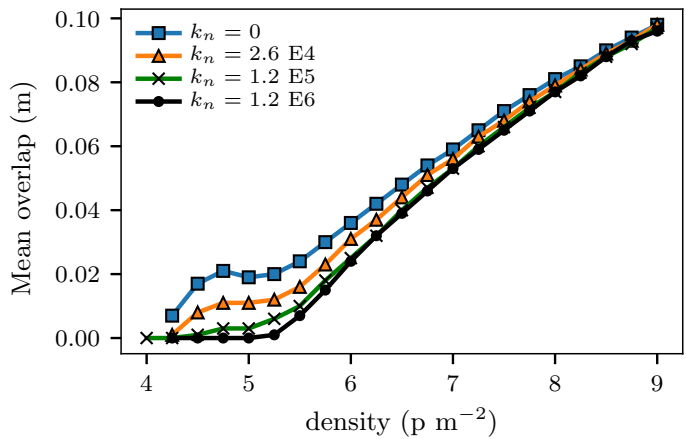}
\caption{\label{overlap_dens} Mean overlap as a function of the 
global density. The global density is the total number of pedestrians per unit 
area. The mean values are averages over all the pedestrians and over time once 
the system reached the stationary state. Both measurements correspond to 
the corridor geometry with desired velocity $v_d=1$. Color 
online only. }
\end{figure}

We can test our hypothesis by computing the mean overlap. 
Fig.~\ref{overlap_dens} shows this magnitude for the corridor situation. Notice 
that ``soft'' pedestrians attain more overlap than the ``stiff'' 
ones, as expected. This is in agreement with Fig.~\ref{overlap_vd} for the 
bottleneck situation (at a fixed value of $v_d$). The curves in 
Fig.~\ref{overlap_vd}, however, do not meet each other as in  
Fig.~\ref{overlap_dens} where all the curves meet each other at
high-density values. This phenomenon occurs due to space limitations
in the corridor that produces overlapping (almost) independently of 
the value of $k_n$. \\

The above results represent an important step in the  investigation.
The inter-pedestrian connectivity differences between the bottleneck geometry
and the corridor geometry were not expected and opens two major questions:  how
do the pedestrians interact with the walls as a function of the body stiffness,
and consequently, how does  this affect the flux across the corridor.\\

\begin{figure}[htbp!]
\centering
\includegraphics[width=0.7\columnwidth]{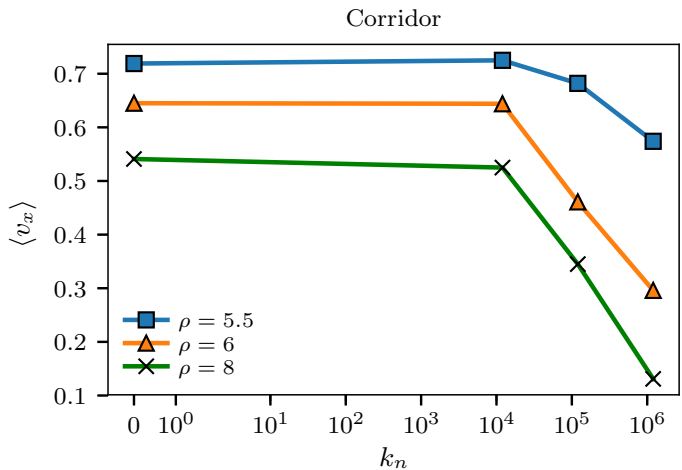}
\caption{\label{kn_vs_vx_corridor} Mean velocity in the longitudinal coordinate
$(v_x)$ as a function of the stiffness $k_n$ for three different global 
densities (see label in the plot). The measurements correspond to a corridor of 
28~m $\times$ 22~m with periodic boundary conditions and $v_d=1$. The average 
was taken along the corridor and along the simulated time. Color online only.}
\end{figure}

We start by computing the mean velocity across the corridor.  
Fig.~\ref{kn_vs_vx_corridor} shows the mean velocity 
$\langle v_x\rangle$ (parallel to the corridor) as a function 
of the stiffness for different global density levels. This 
plot shows a flat pattern for $k_n<10^4$ and a slowing down above this 
threshold. Notice from Fig.~\ref{kn_vs_vx_bottleneck} that this is opposed
to what happens in the bottleneck geometry. \\

The $\langle v_x\rangle$ values in Fig.~\ref{kn_vs_vx_corridor}  were computed
at densities $\rho\geq 5.5$. According to  Fig.~\ref{overlap_dens}, at these
densities, the overlaps among pedestrians attain significant values.
Furthermore,  stiffness values $k_n\geq 10^4$ move the system to a more heavily
connected  stage, and thus, pedestrians have no choice but to walk at  (almost)
the same speed as his (her) neighbors. This  behavior may be envisaged as the
passage from a ``free''  walking movement to a constrained walking movement as
the stiffness increases. A more  physical picture would assimilate the former as
a ``fluid-like  state'' and the latter as a ``solid-like state''. When the
stiffness is  very high (say $k_n=$1.2~E6), all the pedestrians  are expected to
walk at a common velocity. The  pedestrians that walk in physical contact with
the wall are the ones who  determine the velocity of the whole crowd, as
discussed  below.\\

\begin{figure}[!htbp]
\centering
\subfloat[]{\includegraphics[width=0.49\columnwidth]
{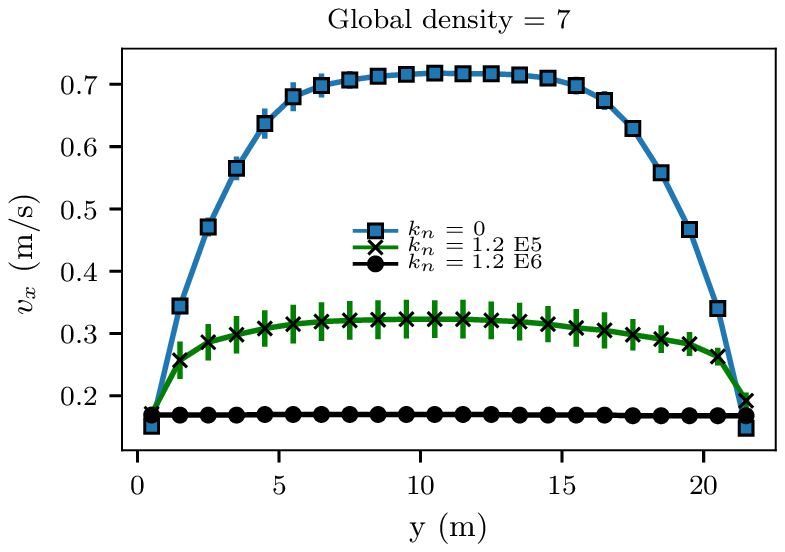}\label{vx_profile}}\ 
\subfloat[]{\includegraphics[width=0.49\columnwidth]
{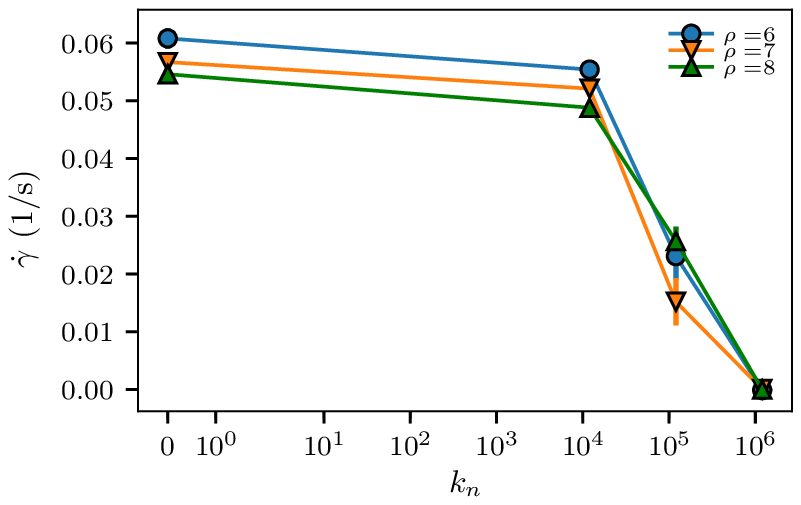}\label{strain_rate_vs_kn}}\\

\caption[width=0.47\columnwidth]{(a) Mean velocity in the $x$ coordinate as a 
function of the transversal coordinate $y$ for three different stiffness values 
$k_n$ (see the label). (b) Strain rate as a function of the stiffness level 
$k_n$ for three different global density values (see label). Data correspond to 
a corridor of 28$\,$m $\times$ 22$\,$m with periodic boundary conditions and 
$v_d=1$. The global density was $\rho=7 p\,$m$^{-2}$. Color online only. }
\label{profile_strain}
\end{figure}

Fig.~\ref{vx_profile} shows the velocity profile ($\langle v_x \rangle$ vs. the
transversal coordinate $y$) for different $k_n$ values. For $k_n=0$, we can see
a parabolic-like velocity profile which means that the friction with the walls
reduces the speed of pedestrians. The velocity profile resembles the Poiseuille
flow (similar to Newtonian and incompressible fluids in a laminar regime). This
behavior was also observed in empirical measurements of crowd dynamics reported
in Ref.~\cite{zhang2013empirical}. But as $k_n$ increases the velocity profile
flattens until becoming (almost) uniform (see Fig.~\ref{vx_profile} for
$k_n=1.2\,$E6). In this scenario, $v_x$ attains a much lower  value  than in the
case of soft pedestrians ($k_n=$0).\\

From the results displayed above, we realize that the crowd behaves like a solid
for very stiff pedestrians. This means that the crowd can not be easily
``deformed''. In this context, deformation means that some parts of the crowd
may be allowed to move faster than other parts of the crowd. \\

The strain rate tensor displays the rate of change of the deformation of a body
in the vicinity of a given point.  We consider the following discrete definition
of the strain rate:

\begin{equation}
\dot{\gamma} = \frac{\langle v_x(\mathrm{center})\rangle
 - \langle v_x(\mathrm{boundary})\rangle }{\left | y(\mathrm{center}) 
 - y(\mathrm{boundary}) \right |} 
\end{equation}

Where $\langle \cdot \rangle$ means the average taken over time. This definition
compares the velocity of the pedestrians close to the wall (boundary) with
respect to the velocity of the pedestrians at the center of the corridor
(center). Thus, the strain rate $\dot{\gamma}$ vanishes (no deformation) as the
stiffness level increases. This phenomenon is shown in
Fig.~\ref{strain_rate_vs_kn} where we can see that the strain rate drops for
high values of $k_n$. \\

We conclude this Section by stressing, once again, that stiffer pedestrians
attain opposite results in corridors with respect to bottlenecks. The walls in
the corridors play a critical role that prevents pedestrians from detaching from
each other. This effect can be observed as an increment in the connectivity
of the contact network and the flattening of the velocity profile (thus reducing
the strain rate). High enough stiffness values stuck the pedestrians
leading to a ``solidification'' of the crowd. All the pedestrians walk at almost
the same velocity at this stage. The pedestrians that are in contact with the
walls are the ones that determine the velocity of the whole crowd. We already
analyzed this situation in Ref~\cite{dorso_2019}. \\

\subsection{\label{Dimensionless}Dimensionless numbers and comparison
 with empirical data}

The final stage of our investigation deals with the plausible values of the
dimensionless numbers $\mathcal{K}_c$ and $\mathcal{K}$, in comparison to the
empirical measurements. We vary the body force coefficient ($k_n$) and the
friction coefficient ($k_t$) to explore different values of $\mathcal{K}_c$ and
$\mathcal{K}$, respectively.\\

\begin{figure}[!htbp]
\centering
\subfloat[]{\includegraphics[width=0.49\columnwidth]
{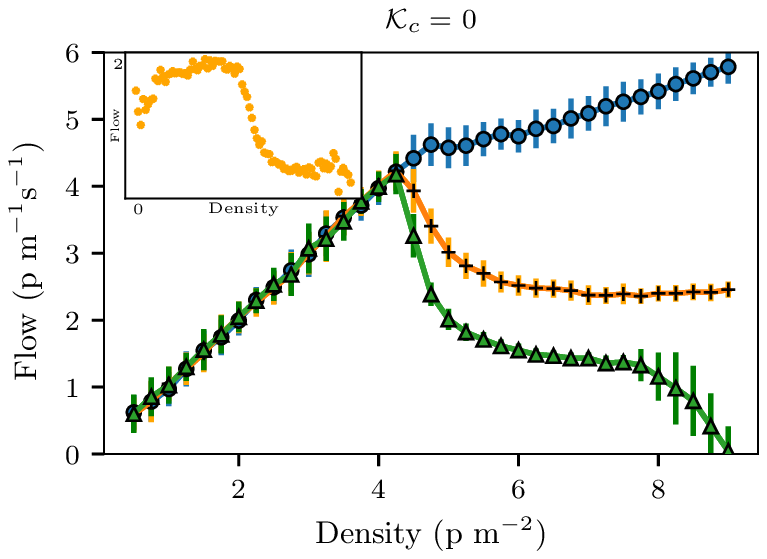}\label{flow_density_no_body_force}}\ 
\subfloat[]{\includegraphics[width=0.49\columnwidth]
{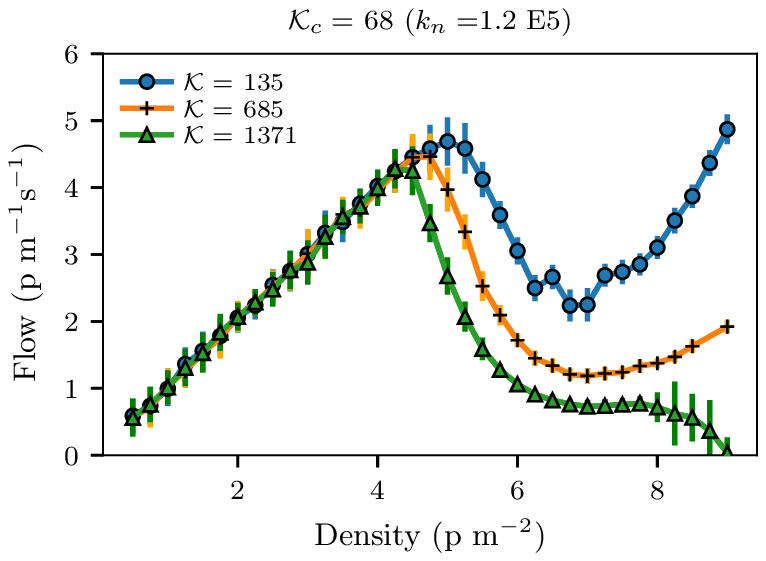}\label{flow_density_body_force}}\\
\caption[width=0.47\columnwidth]{Flow vs. global density. The flow is calculated
in a circular area of $R=1\,$m at the center of the corridor. The circular 
markers correspond to the original friction of the SFM, the "+" symbol 
corresponds to the friction increased by a factor of five and the triangles 
correspond to the friction increased by a factor of ten. The desired velocity 
was $v_d=1$ in all the cases. (a) corresponds to simulations without body force 
($\mathcal{K}_c =$0) and (b) corresponds to a body force with the original value
 of the body stiffness ($\mathcal{K}_c =$68). Color online only.}
\label{flow_density}
\end{figure}

We consider the empirical measurements from Ref.~\cite{helbing_2007}
corresponding to the fundamental diagram obtained at the entrance of the
Jamaraat bridge (see the inset in Fig.~\ref{flow_density_no_body_force}). Our
aim is to reproduce the qualitative behavior of these measurements.\\

Figs. \ref{flow_density} show the pedestrian flow as a function of the global
density (fundamental diagram). Fig.~\ref{flow_density_no_body_force} corresponds
to $\mathcal{K}_c=$ 0 (this is \textit{i.e.} $k_n=0$) while
Fig.~\ref{flow_density_body_force} corresponds to $\mathcal{K}_c=$ 68
corresponding to the original SFM with $k_n=1.2\,$E5. Each curve represents
different friction values (see the caption for details).\\

According to the empirical measurements at Jamaraat, the flow slows down for
high enough densities due to jamming. Notice that the original SFM
(corresponding to $\mathcal{K}=$ 137) does not produce the expected slowing down
for null body force ($\mathcal{K}_c=$ 0). However, when the body force is
present (the original SFM) an ``U'' shape behavior occurs for densities above 5
p$^{-2}$.\\

The increase in $\mathcal{K}$ to $\mathcal{K} = $ 685 (five times the original
SFM value) slows down  the flux for densities above 5 p m$^{-2}$, regardless of
the presence of the body force. Including the body force, however, produces a
subtle increment in the flow for densities higher than 7 p m$^{-2}$ (see orange
curve from Fig.~\ref{flow_density_no_body_force}). \\

A further increment of $\mathcal{K}$ to $\mathcal{K}=$ 1371 (ten times the 
original SFM value) attains a plateau ($\rho>5$ p m$^{-2}$) before vanishing at 
very high densities. This occurs on either $\mathcal{K}_c=$ 0 and 
$\mathcal{K}_c = $ 68.\\

These results suggest that although increasing $\mathcal{K}_c$ slows down the
flux, it is still necessary to increase $\mathcal{K}$ (by increasing the
friction coefficient $k_t$) to avoid an ``U'' shape behavior for extremely high
densities. It is still a challenge to find the optimal dimensionless numbers
$(\mathcal{A},\mathcal{K},\mathcal{K}_c)$ for the fundamental diagram. Our
inspection into the parameter set is a first approach to narrow down the
search.\\

\section{\label{conclusions}Conclusions}

Having adopted as fundamental philosophy of work that the parameters of the SFM
should be properly determined by requiring that the model reproduces
well established experimental data; we explored the effect on the pedestrians
dynamics of the (sometimes neglected) body force in the framework of the SFM. We
showed that the stiffness coefficient ($k_n$) has a significant impact on the
evacuation dynamics (bottleneck) and also in the dynamics of pedestrians walking
along a straight corridor.\\

In the bottleneck geometry, the evacuation time diminishes (pedestrians move
faster) as pedestrians become stiffer along the explored desired velocities.
This phenomenon occurs because stiffer pedestrians reduce the overlapping and
hence the sliding friction intensity. This scenario releases more easily the
pedestrians and reduces the probability of producing a cluster of pedestrians
blocking the exit (blocking cluster). This leads to a more efficient evacuation
dynamics. \\

The opposite behavior is obtained in the corridor geometry with respect to the
bottleneck geometry. The major difference is that pedestrians are limited to the
available space between walls in the corridor geometry.  Walls are not relevant
in the bottleneck geometry. Thus, the overlap between pedestrians is controlled
by the available space. But, stiffer pedestrians are more likely to get stuck.
The whole crowd can be compared to a granular material. Granular materials can
be disordered (amorphous) or ordered depending on how particles interact with
each other. In the present context, at low stiffness levels, the crowd appears
disordered attaining a parabolic velocity profile. If the stiffness level is
high, the whole crowd appears ordered into a lattice (like a crystalline solid)
with a uniform velocity profile that depends on the friction interaction with
the walls.\\

Our efforts to ``tune'' the original SFM to reproduce empirical data (say, the 
fundamental diagram) moved us to explore the dimensionless parameter space.  We 
found that  we can qualitatively reproduce the empirical data if the parameters 
are close to $\mathcal{K}_c < 68$, $\mathcal{K} = 685$ and $\mathcal{A}=14$. 
Nevertheless, this is a first attempt to arrive at suitable parameters, although
 we do not claim these to be optimal. We encourage further research to find 
 dimensionless numbers that may better fit experimental data.\\

\section*{Acknowledgments}
This work was supported by the National Scientific and Technical 
Research Council (spanish: Consejo Nacional de Investigaciones Cient\'\i ficas 
y T\'ecnicas - CONICET, Argentina) grant Programaci\'on Cient\'\i fica 2018 
(UBACYT) Number 20020170100628BA.\\

The authors want to thank the degree students Josefina Catoni and Ayelen Santos
 for providing data acquired at the subway in Buenos Aires, Argentina. \\

G. Frank thanks Universidad Tecnol\'olica Nacional (UTN) for partial
support through Grant PID Number SIUTNBA0006595.

\appendix

\section{\label{appendix1}Reduced-in-units equation of motion}

The SFM description in section~\ref{sfm} introduces seven parameters ($m$, 
$\tau$, $v_d$, $B$, $A$, $k_t$ and $k_n$) attaining for the ``individual'' 
behavior of each pedestrian. The collective dynamic, however, requires a 
smaller set of parameters. In order to identify this smaller set, we introduce 
the following dimensionless magnitudes\\

\begin{equation}
 \left\{\begin{array}{lcl}
         t' & = & t/\tau \\
         r' & = & r/B \\
         v' & = & v/v_d \\
        \end{array}\right.
\end{equation}

The equation of motion (\ref{eqn_motion}) can be rewritten in terms of these 
(dimensionless) magnitudes, while only three (reduced) parameters are needed.\\

\begin{equation}
 \displaystyle\frac{d\mathbf{v}'}{dt'}=
 \hat{\mathbf{e}}_d-\mathbf{v}'+\mathcal{A}\,e^{R'-r'}\,
 \hat{\mathbf{n}}+g(R'-r')\,\bigg[\mathcal{K}\,(\Delta\mathbf{v}'\cdot
 \hat{\mathbf{t}})\,\hat{\mathbf{t}}+\mathcal{K}_c\,\hat{\mathbf{n}}\bigg]
\end{equation}

\noindent where the smaller set ($\mathcal{A}$,$\mathcal{K}$,$\mathcal{K}_c$) 
means\\

\begin{equation}
 \mathcal{A}=\displaystyle\frac{A\,\tau}{m\,v_d}\ \ \ \ , \ \ \ \ 
 \mathcal{K}=\displaystyle\frac{k_t\,B\,\tau}{m}\ \ \ \ , \ \ \ \
 \mathcal{K}_c=\displaystyle\frac{k_n\,B\,\tau}{m\,v_d}
\end{equation}

Notice that the SFM will arrive to similar collective dynamics whenever the 
reduced set ($\mathcal{A}$,$\mathcal{K}$,$\mathcal{K}_c$) remains unchanged 
(although some ``individual'' parameters are allowed to change). For a deep 
explanation on the meaning  of the set 
($\mathcal{A}$,$\mathcal{K}$,$\mathcal{K}_c$) see section~\ref{parameters}. \\




\bibliographystyle{unsrt}
\bibliography{paper}

\end{document}